\documentclass[12pt]{iopart}

\usepackage{graphicx}
\usepackage{bm}
\usepackage{amsbsy}
\usepackage{amssymb}
\usepackage{mathrsfs}

\begin{document}

\title{Mutually unbiased bases: tomography of spin states and star-product scheme}

\author{S N Filippov$^1$ and V I Man'ko$^2$}

\address{$^1$ Moscow Institute of Physics and Technology, Moscow, Russia}

\address{$^2$ P N Lebedev Physical Institute, Moscow, Russia}

\eads{\mailto{sergey.filippov@phystech.edu},
\mailto{manko@sci.lebedev.ru}}

\begin{abstract}
Mutually unbiased bases (MUBs) are considered within the framework
of a generic star-product scheme. We rederive that a full set of
MUBs is adequate for a spin tomography, i.e. knowledge of all
probabilities to find a system in each MUB-state is enough for a
state reconstruction. Extending the ideas of the
tomographic-probability representation and the star-product scheme
to MUB-tomography, dequantizer and quantizer operators for
MUB-symbols of spin states and operators are introduced, ordinary
and dual star-product kernels are found. Since MUB-projectors are
to obey specific rules of the star-product scheme, we reveal the
Lie algebraic structure of MUB-projectors and derive new relations
on triple- and four-products of MUB-projectors. Example of qubits
is considered in detail. MUB-tomography by means of Stern-Gerlach
apparatus is discussed.
\end{abstract}


\maketitle

\section{\label{introduction}Introduction}
Since the early days of quantum mechanics, much attention has been
paid to a problem of a good description of quantum states. The
notions of wave function $\psi$ and density matrix $\rho$ are most
widely known and used. Nevertheless, these notions give rise to a
problem of interpretation, especially, in case of measuring a
quantum state. Outcomes of quantum observables are known to be
probabilistic. In view of this, quasiprobability distribution
functions like Wigner $W$-function~\cite{wigner},
Sudarshan-Glauber $P$-function~\cite{sudarshan,glauber}, and
Husimi $Q$-function~\cite{husimi} are often used in quantum optics
along with the wave function and density matrix formalism. The
main drawback of $W$, $P$, and $Q$ functions is that they cannot
be measured experimentally. The problem of measuring quantum
states resulted in developing quantum tomography, and then in a
tomographic-probability representation of quantum mechanics
(historical background is given in the review~\cite{ibort}).
According to such a representation, any quantum state of light is
described by measurable tomograms: optical, symplectic, and
photon-number ones (see, e.g. the review~\cite{oman'ko-97}). As
far as a finite dimensional Hilbert space is concerned, one can
alternatively utilize spin tomogram~\cite{dodonovPLA,oman'ko-jetp}
and spin tomogram with a finite number of
rotations~\cite{serg-inverse-spin}. Apart from being appropriate
for reconstructing the density matrix, quantum tomograms are
themselves notions of quantum states. Within the framework of the
tomographic-probability representation, operators are described by
tomographic symbols satisfying rules of the corresponding
star-product scheme. We cannot help mentioning some of the
probability-based approaches to quantum mechanics, namely, the
expectation-value
representation~\cite{weigert-PRL,weigert-mic-povm} and the
Bayesian interpretation~\cite{fuchs-perimeter,ericsson} utilizing
a symmetric informationally complete positive operator-valued
measure (SIC-POVMs are discussed, e.g.,
in~\cite{caves,renes,fuchs-2010}).

The aim of this paper is to develop the
star-product~\cite{stratonovich} quantization scheme based on
mutually unbiased bases (MUBs)~\cite{ivanovic,wooters}. MUBs
represent themselves a highly symmetrical structure and have many
interesting properties (see, e.g.~\cite{klappenecker-rotteler} and
references therein). For example, a full set of MUBs is known to
exist whenever the dimension of Hilbert space is a prime number or
the power of a prime. We consider neither the problem of existence
of MUBs in a given Hilbert space nor the problem how many MUBs
there exist. We assume that the full set of MUBs is known for the
space involved. In this article, we combine MUBs with the
tomographic-probability representation. As a result,
MUB-tomography of spin states is introduced, MUB-symbols of
quantum operators are considered within the framework of the
star-product scheme, the Lie algebraic structure of MUBs is
pointed out, and new properties on MUB-projectors are derived.
Special attention is focused on qubits.

The paper is organized as follows.

In \Sref{section-MUBs}, MUBs are briefly reviewed. In
\Sref{section-MUB-tomography}, MUB-based tomography is considered,
scanning and reconstruction procedures are presented. In
\Sref{section-star-product}, we follow the ideas of a generic
star-product scheme \cite{oman'ko-JPA,oman'ko-vitale} and analyze
the star product of MUB-symbols. In \Sref{section-qubits}, an
example of qubits is considered in detail. In \Sref{section-S-G},
a practical realization of MUB-tomography by means of
Stern-Gerlach apparatus is discussed. In \Sref{conclusions},
conclusions and prospects are presented.

\section{\label{section-MUBs}Mutually unbiased bases}
Let us consider a $d$-dimensional Hilbert space endowed with a
full set of mutually unbiased bases. If this is the case, MUBs
consist of $d+1$ bases $\{|a\alpha\rangle\}_{\alpha=0}^{d-1}$,
where $a=0,\ldots,d$ is responsible for the basis number, an index
$\alpha=0,\ldots,d-1$ refers to one of the basis states belonging
to the particular basis $a$. MUBs are to satisfy the following
property:
\begin{equation}
\label{MUBs} \left| \langle a\alpha | b\beta \rangle \right|^2 =
\frac{1}{d}(1-\delta_{a,b}) + \delta_{a,b}\delta_{\alpha,\beta},
\end{equation}
\noindent where $\delta_{a,b}$ is a Kronecker delta symbol. Eq.
(\ref{MUBs}) implies that each basis is orthonormal and arbitrary
two states belonging to different bases are equiangular, i.e.
$\left| \langle a\alpha | b\beta \rangle \right|^2 = \frac{1}{d}$
if $a \ne b$.

Let us now consider rank-1 MUB-projectors $\hat{\Pi}_{a\alpha} =
|a\alpha\rangle \langle a\alpha |$.

Obviously, operators $\hat{\Pi}_{a\alpha}$ are semi-positive and
satisfy the trace relation ${\rm Tr}\big[ \hat{\Pi}_{a\alpha}
\hat{\Pi}_{b\beta} \big] = \frac{1}{d}(1-\delta_{a,b}) +
\delta_{a,b}\delta_{\alpha,\beta}$. An immediate consequence of
orthonormality is
\begin{eqnarray}
\label{sum-alpha} \sum_{\alpha=0}^{d-1} \hat{\Pi}_{a\alpha} =
\hat{I} {\rm ~~ for ~
all ~} a = 0,\ldots,d, \\
\label{sum-a-alpha} \sum_{a=0}^{d} \sum_{\alpha=0}^{d-1}
\hat{\Pi}_{a\alpha} = (d+1) \hat{I},
\end{eqnarray}
\noindent where $\hat{I}$ is the identity operator. Since the
relation (\ref{sum-alpha}) is valid for all $a = 0,\ldots,d$, the
total number of linearly independent operators
$\hat{\Pi}_{a\alpha}$ equals $1+(d+1)(d-1) = d^2$. This means that
the identity operator $\hat{I}$ together with $d^2-1$ operators
$\{\hat{\Pi}_{a\alpha}\}$, $a=0,\ldots,d$, $\alpha = 0,\ldots,d-2$
form a basis in $d$-dimensional Hilbert space. As a result, any
operator including the density operator $\hat{\rho}$ of a quantum
state can be resolved through these basis operators. Indeed,
\begin{equation}
\label{rho-expansion} \hat{\rho} = c_{I}\hat{I} +
\sum_{b=0}^{d}\sum_{\beta=0}^{d-2} c_{b\beta} \hat{\Pi}_{b\beta},
\end{equation}
\noindent where $c_I$ and $c_{b\beta}$ are real parameters because
both $\hat{\rho}$ and $\hat{\Pi}_{b\beta}$ are Hermitian. Applying
the trace operation to both sides of (\ref{rho-expansion}) and
utilizing ${\rm Tr}\hat{\rho} = {\rm Tr} \hat{\Pi}_{b\beta} = 1$,
we readily obtain
\begin{eqnarray}
c_I = \frac{1}{d}\left( 1 - \sum_{b=0}^{d}\sum_{\beta=0}^{d-2}
c_{b\beta} \right), \\
\label{rho-expansion-v2} \hat{\rho} = \frac{1}{d}\hat{I} +
\sum_{b=0}^{d}\sum_{\beta=0}^{d-2} c_{b\beta} \left(
\hat{\Pi}_{b\beta} - \frac{1}{d}\hat{I} \right).
\end{eqnarray}
Taking into account the non-negativity of operators
$\hat{\Pi}_{a\alpha}$ and the sum rule (\ref{sum-a-alpha}), it is
not hard to see that operators $\hat{E}_{a\alpha} =
(d+1)^{-1}\hat{\Pi}_{a\alpha}$ altogether form a positive
operator-valued measure (POVM). We will refer to such a POVM as
MUB-POVM.

\section{\label{section-MUB-tomography}MUB-tomography}
In an experiment, probabilities of measurement outcomes are only
accessible. Tomography is a procedure allowing one to reconstruct
density operator $\hat{\rho}$ with the help of measured
probabilities. We will consider projective (von Neumann)
measurements associated with MUBs. In other words, we assume that
the probabilities
\begin{equation}
\label{p-TrRhoPi} p_{a\alpha} = \langle a\alpha | \hat{\rho} |
a\alpha \rangle = {\rm Tr} \big[ \hat{\rho} \hat{\Pi}_{a\alpha}
\big]
\end{equation}
\noindent are known for all $a=0,\ldots,d$, $\alpha=0,\ldots,d-1$.
As a consequence of expressions
(\ref{sum-alpha})--(\ref{sum-a-alpha}) we obtain the following
normalization conditions:
\begin{equation}
\label{p-normalization} \sum_{\alpha=0}^{d-1} p_{a\alpha} = 1,
\quad \sum_{a=0}^{d} \sum_{\alpha=0}^{d-1} p_{a\alpha} = d+1.
\end{equation}

The physical meaning of $p_{a\alpha}$ is the probability to find a
system in the state $|a\alpha\rangle$ which is itself an element
of MUBs. The problem is to express the density operator
$\hat{\rho}$ through probabilities $p_{a\alpha}$. This problem is
solved in~\cite{wooters-fields}. For the sake of the subsequent
consideration we rederive the result and present it in a slightly
different manner.

\textbf{Proposition}. A reconstruction procedure of MUB-tomography
reads
\begin{equation}
\label{proposition} \hat{\rho} = \sum_{b=0}^{d}
\sum_{\beta=0}^{d-1} p_{b\beta} \left( \hat{\Pi}_{b\beta} -
\frac{1}{d+1} \hat{I} \right).
\end{equation}

\textit{Proof}. Multiplying both sides of Eq.
(\ref{rho-expansion-v2}) by $\hat{\Pi}_{a\alpha}$ and applying the
trace operation, we obtain
\begin{equation}
\label{p-c} p_{a\alpha} = \frac{1}{d} + \sum_{b=0}^{d}
\sum_{\beta=0}^{d-2} M_{a\alpha,b\beta} c_{b\beta},
\end{equation}
\noindent where $M_{a\alpha,b\beta} = {\rm Tr} \big[
\hat{\Pi}_{a\alpha} \hat{\Pi}_{b\beta} \big] - \frac{1}{d} =
\delta_{a,b} (\delta_{\alpha,\beta} - \frac{1}{d})$. The composed
index $a\alpha$ can be considered as a single one: $k=ad+\alpha$,
$k=0,\ldots,d^2+d-1$. Then formula (\ref{p-c}) is nothing else as
a linear system of equations with respect to $c_k$, and can be
rewritten in matrix form as follows:
\begin{equation}
\left(%
\begin{array}{c}
  p_0 - 1/d\\
  \vdots \\
  p_{d^2+d-1} -1/d \\
\end{array}%
\right) = M \left(%
\begin{array}{c}
  c_0\\
  \vdots \\
  c_{d^2+d-1} \\
\end{array}%
\right),
\end{equation}
\noindent where $M$ is a $(d^2-1) \times (d^2-1)$ block-diagonal
matrix of the form
\begin{equation} \label{M} M = \left(
\begin{array}{cccc}
\mathcal{M} & \multicolumn{3}{|c}{}\\
\cline{1-2} \multicolumn{1}{c|}{} & \mathcal{M}
&\multicolumn{2}{|c}
{\raisebox{1.5ex}[0pt]{\parbox{12pt}{\Huge 0}}}\\
\cline{2-2} \multicolumn{2}{c}{}& \ddots &   \\
\cline{4-4} \multicolumn{3}{c|}
{\raisebox{1.5ex}[0pt]{\parbox{12pt}{\Huge 0}}} & \mathcal{M}
\end{array}
 \right),
\end{equation}
\noindent with each $(d-1) \times (d-1)$ block $\mathcal{M}$ being
equal to
\begin{equation}
\mathcal{M} = \left(%
\begin{array}{cccc}
  1-1/d & -1/d & \cdots & -1/d \\
  -1/d & 1-1/d & \cdots & -1/d \\
  \vdots & \vdots & \ddots & \vdots \\
  -1/d & -1/d & \cdots & 1-1/d \\
\end{array}%
\right).
\end{equation}

A direct calculation of the inverse matrix yields
\begin{equation} \label{M-inverse} \mathcal{M}^{-1} = \left(%
\begin{array}{cccc}
  2 & 1 & \cdots & 1 \\
  1 & 2 & \cdots & 1 \\
  \vdots & \vdots & \ddots & \vdots \\
  1 & 1 & \cdots & 2 \\
\end{array}%
\right),
\end{equation}
\noindent from which it is not hard to find an explicit solution
of system (\ref{p-c}). The result is
\begin{equation}
\label{c-through-p} c_{b\beta} = p_{b\beta} - \frac{1}{d} +
\sum_{\widetilde{\beta}=0}^{d-2} \left( p_{b\widetilde{\beta}} -
\frac{1}{d} \right) = p_{b\beta} - p_{b,d-1}.
\end{equation}
\noindent The right-hand equality in (\ref{c-through-p}) is due to
normalization condition (\ref{p-normalization}). Substituting the
obtained value of $c_{b\beta}$ in (\ref{rho-expansion-v2}) and
making use of (\ref{sum-a-alpha}) and (\ref{p-normalization}), we
finally have
\begin{eqnarray}
\hat{\rho} &=& \frac{1}{d(d+1)} \sum_{b=0}^{d}
\sum_{\beta=0}^{d-1} p_{b\beta} \hat{I} + \sum_{b=0}^{d}
\sum_{\beta=0}^{d-2} (p_{b\beta}-p_{b,d-1}) \left(
\hat{\Pi}_{b\beta} - \frac{1}{d} \hat{I} \right) \nonumber\\
&=& \sum_{b=0}^{d} \sum_{\beta=0}^{d-2} p_{b\beta} \left(
\hat{\Pi}_{b\beta} - \frac{1}{d+1} \hat{I} \right) -
\sum_{b=0}^{d} p_{b,d-1} \sum_{\beta=0}^{d-2} \left(
\hat{\Pi}_{b\beta} - \frac{1}{d} \hat{I} \right) \nonumber\\
&& + \sum_{b=0}^{d} p_{b,d-1} \frac{1}{d(d+1)} \hat{I}
\nonumber\\
&=& \sum_{b=0}^{d} \sum_{\beta=0}^{d-2} p_{b\beta} \left(
\hat{\Pi}_{b\beta} - \frac{1}{d+1} \hat{I} \right) -
\sum_{b=0}^{d} p_{b,d-1} \left( \frac{1}{d+1} \hat{I} -
\hat{\Pi}_{b,d-1} \right) \nonumber\\
&=& \sum_{b=0}^{d} \sum_{\beta=0}^{d-1} p_{b\beta} \left(
\hat{\Pi}_{b\beta} - \frac{1}{d+1} \hat{I} \right).
\end{eqnarray}
This completes the proof of proposition. $\blacksquare$

Note that reconstruction formula (\ref{proposition}) is not unique
and many alternative expressions can be found, but we will use
this formula in view of its symmetry. Also, an immediate
consequence of formula (\ref{proposition}) is that if MUBs exist
then any qudit state can be represented by a single probability
distribution $\{(d+1)^{-1}p_{a\alpha}\}$. From this point of view,
MUB-based representation of qudit states is a partial case of an
inverse spin-$s$ portrait method~\cite{serg-inverse-spin} with an
extra requirement on the symmetry.

\section{\label{section-star-product}Star-product scheme}
Any operator $\hat{A}$ acting on Hilbert space of quantum states
can be alternatively described by a symbol $f_{A}({\bf x})$ which
is a function of a particular set of variables ${\bf x}$. The
relation between $\hat{A}$ and $f_A({\bf x})$ is defined through
\begin{equation}
\label{symbol} f_A({\bf x}) = {\rm Tr} \big[ \hat{A} \hat{U}({\bf
x}) \big], \quad \hat{A} = \int d{\bf x} f_A({\bf x}) \hat{D}({\bf
x}),
\end{equation}
\noindent where $\hat{U}({\bf x})$ and $\hat{D}({\bf x})$ are
dequantizer and quantizer operators, respectively; an explicit
form of the sign of integration $\int d{\bf x}$ depends on the
scheme used. It is worth noting that substituting the second
equality (\ref{symbol}) for $\hat{A}$ in the definition of symbol
of operator, we readily obtain that a function $\mathfrak{D}({\bf
x}_1,{\bf x}) = {\rm Tr} \big[ \hat{D}({\bf x}_1) \hat{U}({\bf x})
\big]$ has a sense of delta-function on symbols, i.e. $\int d{\bf
x}_1 \mathfrak{D}({\bf x}_1,{\bf x}) f_A({\bf x}_1) = f_A({\bf
x})$.

Using such a formalism, one can deal with symbols of operators
instead of operators themselves. The rules of addition and
multiplication for operators are then transformed into the
following rules for symbols:
\begin{eqnarray}
& \hat{C} = \hat{A} + \hat{B}: & \quad f_C({\bf x}) = f_A({\bf x})
+ f_B({\bf x}),\\
& \hat{C} = c\hat{A}: & \quad f_C({\bf x}) = c f_A({\bf x}),\\
& \hat{C} = \hat{A} \hat{B}: & \quad f_C({\bf x}) = (f_A \star
f_B) ({\bf x}),
\end{eqnarray}
\noindent where by star we denote a star product of symbols
defined through
\begin{eqnarray}
&& (f_A \star f_B) ({\bf x}) = \int \!\!\! \int d{\bf x}_1 d{\bf
x}_2 f_A({\bf
x}_1) f_B({\bf x}_2) K({\bf x}_1,{\bf x}_2,{\bf x}),\\
&& K({\bf x}_1,{\bf x}_2,{\bf x}) = {\rm Tr} \big[ \hat{D}({\bf
x}_1) \hat{D}({\bf x}_2) \hat{U}({\bf x}) \big].
\end{eqnarray}
\noindent The term $K({\bf x}_1,{\bf x}_2,{\bf x})$ is usually
referred to as star-product kernel
\cite{oman'ko-JPA,oman'ko-vitale}. As the star product is
associative, from the relation $f_A \star f_B \star f_C = f_A
\star (f_B \star f_C) = (f_A \star f_B) \star f_C$ we readily
obtain a star-product kernel of 3 symbols $K^{(3)}({\bf x}_1,{\bf
x}_2,{\bf x}_3,{\bf x})$ as well as an additional requirement on
the star-product kernel
\begin{eqnarray}
\label{KK-KK} K^{(3)}({\bf x}_1,{\bf x}_2,{\bf x}_3,{\bf x}) &=&
\int d{\bf y} K({\bf x}_1,{\bf x}_2,{\bf y}) K({\bf y},{\bf
x}_3,{\bf x}) \nonumber\\
&=& \int d{\bf y} K({\bf x}_1,{\bf y},{\bf x}) K({\bf x}_2,{\bf
x}_3,{\bf y}),
\end{eqnarray}
\noindent which is valid for all sets of variables ${\bf x}_1$,
${\bf x}_2$, ${\bf x}_3$, and ${\bf x}$.

\subsection{Dual star-product scheme}
The star-product scheme of the form
\begin{equation}
\label{dual-symbol} f_A^d({\bf x}) = {\rm Tr} \big[ \hat{A}
\hat{D}({\bf x}) \big], \quad \hat{A} = \int d{\bf x} f_A^d({\bf
x}) \hat{U}({\bf x})
\end{equation}
\noindent is called dual with respect to (\ref{symbol}).
Apparently, dual star-product kernel reads $K^d({\bf x}_1,{\bf
x}_2,{\bf x}) = {\rm Tr} \big[ \hat{U}({\bf x}_1) \hat{U}({\bf
x}_2) \hat{D}({\bf x}) \big]$ and satisfies the relation
(\ref{KK-KK}).

\subsection{\label{section-intertwining}Relation between
tomographic schemes} Suppose that we are given two star-product
schemes: (i) $\hat{U}({\bf x})$, $\hat{D}({\bf x})$ and (ii)
$\hat{\mathcal{U}}(\boldsymbol{\xi})$,
$\hat{\mathcal{D}}(\boldsymbol{\xi})$. A relation between the
corresponding symbols is
\begin{equation}
f_{\rm i}({\bf x}) = \int d{\boldsymbol{\xi}} f_{\rm
ii}(\boldsymbol{\xi}) K_{\rm ii \rightarrow
i}(\boldsymbol{\xi},{\bf x}), \quad f_{\rm ii}(\boldsymbol{\xi}) =
\int d{\bf x} f_{\rm i}({\bf x}) K_{\rm i \rightarrow ii}({\bf
x},\boldsymbol{\xi}),
\end{equation}
\noindent where intertwining kernels $K_{\rm ii \rightarrow
i}(\boldsymbol{\xi},{\bf x})$ and $K_{\rm i \rightarrow ii}({\bf
x},\boldsymbol{\xi})$ read
\begin{equation}
\label{intertwining-kernels} K_{\rm ii \rightarrow
i}(\boldsymbol{\xi},{\bf x}) = {\rm Tr} \big[
\hat{\mathcal{D}}(\boldsymbol{\xi}) \hat{U}({\bf x}) \big], \quad
K_{\rm i \rightarrow ii}({\bf x},\boldsymbol{\xi}) = {\rm Tr}
\big[ \hat{D}({\bf x}) \hat{\mathcal{U}}(\boldsymbol{\xi}) \big].
\end{equation}

The relation between MUB-symbols and symbols of a symmetric
informationally complete POVM (SIC-POVM) will be considered for
qubits in \Sref{section-qubits}.

\subsection{MUB star-product scheme}
Comparing MUB-scanning procedure (\ref{p-TrRhoPi}) and
reconstruction procedure (\ref{proposition}) with the scheme
(\ref{symbol}), it is not hard to see that MUB-tomography can be
treated as a star-product scheme with the following dequantizer
and quantizer operators:
\begin{equation}
\hat{U}_{a\alpha} = \hat{\Pi}_{a\alpha}, \quad \hat{D}_{a\alpha} =
\hat{\Pi}_{a\alpha} - \frac{1}{d+1}\hat{I},
\end{equation}
\noindent where ${\bf x} = \{ a, \alpha \}$, $a=0,\ldots,d$,
$\alpha=0,\ldots,d-1$ and $\int d{\bf x} = \sum_{a=0}^{d}
\sum_{\alpha=0}^{d-1}$, i.e. ${\bf x}$ is a set of discrete
variables and integration $\int d{\bf x}$ implies summation.
MUB-tomographic symbol of any operator $\hat{A}$ acting on
$d$-dimensional Hilbert space is
\begin{eqnarray}
\label{MUB-symbol} && f_A(a,\alpha) = {\rm Tr} \big[ \hat{A}
\hat{\Pi}_{a\alpha}
\big], \\
&& \label{MUB-A-reconstruct} \hat{A} = \sum_{a=0}^{d}
\sum_{\alpha=0}^{d-1} f_A(a,\alpha) \left( \hat{\Pi}_{a\alpha} -
\frac{1}{d+1}\hat{I} \right).
\end{eqnarray}

Delta-function on MUB-tomographic symbols is
\begin{eqnarray}
\mathfrak{D}(a,\alpha;b,\beta) &=& {\rm Tr} \big[
\hat{D}_{a\alpha} \hat{U}_{b\beta} \big] = {\rm Tr} \big[
\hat{\Pi}_{a\alpha}
\hat{\Pi}_{b\beta} \big] - \frac{1}{d+1} \nonumber\\
&=& \frac{1}{d(d+1)} +
\delta_{a,b}\left(\delta_{\alpha,\beta}-\frac{1}{d}\right).
\end{eqnarray}

\noindent Note that the obtained delta-function contains extra
terms in addition to the Kronecker delta symbol
$\delta_{a,b}\delta_{\alpha,\beta}$. There is no contradiction
here and it can easily be checked that the residual part always
gives zero while summation with any MUB-symbol.

MUB star-product kernel is expressed through MUB triple product
$T_{a\alpha,b\beta,c\gamma} = {\rm Tr} \big[ \hat{\Pi}_{a\alpha}
\hat{\Pi}_{b\beta} \hat{\Pi}_{c\gamma} \big]$ as follows:
\begin{eqnarray}
\label{kernel} K(a,\alpha;b,\beta;c,\gamma) = {\rm Tr} \big[
\hat{D}_{a\alpha}
\hat{D}_{b\beta} \hat{U}_{c\gamma} \big] \nonumber\\
= T_{a\alpha,b\beta,c\gamma} +
\frac{\delta_{a,c}+\delta_{b,c}}{d(d+1)} -
\frac{\delta_{a,c}\delta_{\alpha,\gamma} +
\delta_{b,c}\delta_{\beta,\gamma}}{d+1} - \frac{d+2}{d(d+1)^2}.
\end{eqnarray}

Star-product kernel $K(a,\alpha;b,\beta;c,\gamma)$ necessarily
meets the condition (\ref{KK-KK}), from which we derive a new
relation on MUB triple product
\begin{eqnarray}
&& \sum_{c=0}^{d} \sum_{\gamma=0}^{d-1} \left(
T_{a\alpha,b\beta,c\gamma} T_{c\gamma,k\varkappa,l\lambda} -
T_{a\alpha,c\gamma,l\lambda} T_{b\beta,k\varkappa,c\gamma} \right)
\nonumber\\
&& = \left( \frac{1}{d}(1-\delta_{a,b}) +
\delta_{a,b}\delta_{\alpha,\beta} \right) \left(
\frac{1}{d}(1-\delta_{k,l}) +
\delta_{k,l}\delta_{\varkappa,\lambda} \right) \nonumber\\
&& - \left( \frac{1}{d}(1-\delta_{a,l}) +
\delta_{a,l}\delta_{\alpha,\lambda} \right) \left(
\frac{1}{d}(1-\delta_{b,k}) + \delta_{b,k}\delta_{\beta,\varkappa}
\right)
\end{eqnarray}
\noindent and find an expression which relates the four-product
${\rm Tr} \big[ \hat{\Pi}_{a\alpha} \hat{\Pi}_{b\beta}
\hat{\Pi}_{k\varkappa} \hat{\Pi}_{l\lambda} \big]$ and triple
product
\begin{eqnarray}
\label{four-product} && {\rm Tr} \big[ \hat{\Pi}_{a\alpha}
\hat{\Pi}_{b\beta} \hat{\Pi}_{k\varkappa} \hat{\Pi}_{l\lambda}
\big] = \sum_{c=0}^{d} \sum_{\gamma=0}^{d-1}
T_{a\alpha,b\beta,c\gamma}
T_{c\gamma,k\varkappa,l\lambda} \nonumber\\
&& - \left( \frac{1}{d}(1-\delta_{a,b}) +
\delta_{a,b}\delta_{\alpha,\beta} \right) \left(
\frac{1}{d}(1-\delta_{k,l}) +
\delta_{k,l}\delta_{\varkappa,\lambda} \right).
\end{eqnarray}

It is worth mentioning that the same result can be alternatively
obtained by using dual MUB star-product kernel of the form
\begin{eqnarray}
\label{kernel-dual} K^d(a,\alpha;b,\beta;c,\gamma) = {\rm Tr}
\big[ \hat{D}_{a\alpha}
\hat{D}_{b\beta} \hat{U}_{c\gamma} \big] \nonumber\\
= T_{a\alpha,b\beta,c\gamma} - \frac{1}{d+1} \left( \frac{1}{d}
(1-\delta_{a,b}) + \delta_{a,b}\delta_{\alpha,\beta} \right).
\end{eqnarray}

\subsection{Lie algebraic structure of MUB-POVM}
The developed MUB star-product scheme enables to reveal the Lie
algebraic structure of MUB-projectors. In fact, following the
ideas of \cite{fuchs-2010}, let us consider a commutator $\hat{C}
= \left[ \hat{\Pi}_{a\alpha}, \hat{\Pi}_{b\beta} \right] =
\hat{\Pi}_{a\alpha}\hat{\Pi}_{b\beta} -
\hat{\Pi}_{b\beta}\hat{\Pi}_{a\alpha}$. Since MUB-projectors are
Hermitian, we obtain $\hat{C}^{\dag} = - \hat{C}$. This means that
the MUB-symbol of such a commutator is purely imaginary, that is
\begin{equation}
f_{C}(c,\gamma) = {\rm Tr} \big[ \hat{C} \hat{\Pi}_{c\gamma} \big]
= T_{a\alpha,b\beta,c\gamma} - T_{b\beta,a\alpha,c\gamma} = i
J_{a\alpha,b\beta,c\gamma},
\end{equation}
\noindent where $J_{a\alpha,b\beta,c\gamma}$ is real and satisfies
the condition $\sum_{\gamma=0}^{d-1} J_{a\alpha,b\beta,c\gamma} =
0$. Using this condition and reconstruction formula
(\ref{MUB-A-reconstruct}), we readily obtain
\begin{eqnarray}
\hat{C} = \sum_{c=0}^{d} \sum_{\gamma=0}^{d-1} i
J_{a\alpha,b\beta,c\gamma} \left( \hat{\Pi}_{c\gamma} -
\frac{1}{d+1} \hat{I} \right), \\
\left[ \hat{\Pi}_{a\alpha}, \hat{\Pi}_{b\beta} \right] =
\sum_{c=0}^{d} \sum_{\gamma=0}^{d-1} i J_{a\alpha,b\beta,c\gamma}
\hat{\Pi}_{c\gamma}.
\end{eqnarray}
\noindent The latter equation means that MUB-projectors form the
Lie algebra ${\rm gl}(d,\mathbb{C})$, with
$iJ_{a\alpha,b\beta,c\gamma}$ being structure constants.
Evidently, MUB-POVM effects $\{\hat{E}_{a\alpha}\}$ satisfy
$\left[ \hat{E}_{a\alpha}, \hat{E}_{b\beta} \right] =
(d+1)^{-1}\sum_{c=0}^{d} \sum_{\gamma=0}^{d-1} i
J_{a\alpha,b\beta,c\gamma} \hat{E}_{c\gamma}$.

\section{\label{section-qubits} MUB star-product scheme for qubits}
MUB-projectors in 2-dimensional Hilbert space can be chosen as
follows:
\begin{eqnarray}
\hat{\Pi}_{a=0,\alpha=0} = \frac{1}{2}(\hat{I}+\hat{\sigma}_x),
\quad \hat{\Pi}_{a=0,\alpha=1} =
\frac{1}{2}(\hat{I}-\hat{\sigma}_x),\\
\hat{\Pi}_{a=1,\alpha=0} = \frac{1}{2}(\hat{I}+\hat{\sigma}_y),
\quad \hat{\Pi}_{a=1,\alpha=1} =
\frac{1}{2}(\hat{I}-\hat{\sigma}_y),\\
\hat{\Pi}_{a=2,\alpha=0} = \frac{1}{2}(\hat{I}+\hat{\sigma}_z),
\quad \hat{\Pi}_{a=2,\alpha=1} =
\frac{1}{2}(\hat{I}-\hat{\sigma}_z),
\end{eqnarray}
\noindent where
$\hat{\boldsymbol{\sigma}}=(\hat{\sigma}_x,\hat{\sigma}_y,\hat{\sigma}_z)$
is a set of Pauli operators.

\noindent The delta-function on MUB symbols for qubits is
$\mathfrak{D}(a,\alpha;b,\beta) = \frac{1}{6} +
\delta_{a,b}(\delta_{\alpha,\beta}-\frac{1}{2})$.

\noindent The direct calculation shows that the triple product of
MUB-projectors for qubits reads
\begin{eqnarray}
T_{a\alpha,b\beta,c\gamma} &=& \frac{1}{4} \Big[ 1 +
2(\delta_{a,b}\delta_{\alpha,\beta} +
\delta_{b,c}\delta_{\beta,\gamma} +
\delta_{c,a}\delta_{\gamma,\alpha}) \nonumber\\
&& - (\delta_{a,b} + \delta_{b,c} + \delta_{c,a}) \nonumber\\
&& + i \varepsilon_{abc} (\delta_{\alpha,0}-\delta_{\alpha,1})
(\delta_{\beta,0}-\delta_{\beta,1})
(\delta_{\gamma,0}-\delta_{\gamma,1}) \Big],
\end{eqnarray}
\noindent where $\varepsilon_{abc}$ is Levi-Civita symbol.
Substituting the obtained triple product in formulas
(\ref{kernel}), (\ref{kernel-dual}), and (\ref{four-product}), it
is easy to calculate the ordinary and dual MUB star-product
kernels as well as the four-product for qubits.

\subsection{Relation to SIC star-product scheme}
A star-product scheme based on symmetric informationally complete
POVM is considered in~\cite{serg-sic}. A self-dual star-product
scheme that is very similar to SIC star-product scheme is
considered in~\cite{livine}. In case of qubits, SIC-projectors
read $\hat{\mathcal{P}}_{k} = \frac{1}{2}
(\hat{I}+(\hat{\boldsymbol{\sigma}} \cdot {\bf n}_{k}))$,
$k=1,\ldots,4$, where ${\bf n}_1 = \frac{1}{\sqrt{3}}(1,1,1)$,
${\bf n}_2 = \frac{1}{\sqrt{3}}(1,-1,-1)$, ${\bf n}_3 =
\frac{1}{\sqrt{3}}(-1,1,-1)$, and ${\bf n}_4 =
\frac{1}{\sqrt{3}}(-1,-1,1)$. The dequantizer is
$\hat{\mathcal{U}}_k = \frac{1}{2} \hat{\mathcal{P}}_{k}$ and the
quantizer is $\hat{\mathcal{D}}_k = 3\hat{\mathcal{P}}_{k} -
\hat{I}$. A calculation of intertwining kernels
(\ref{intertwining-kernels}) between MUB and SIC star-product
schemes yields
\begin{eqnarray}
&& K_{\rm SIC \rightarrow MUB} (k; a,\alpha) = \frac{1}{2} \left(1
+ \sqrt{3} S(k; a,\alpha) \right), \\
&& K_{\rm MUB \rightarrow SIC} (a,\alpha; k) = \frac{1}{12}
\left(1 + \sqrt{3} S(k; a,\alpha) \right),
\end{eqnarray}
\noindent where $S(k; a,\alpha)$ is a sign function taking values
$\pm 1$ in accordance with table 1. We hope that analogues simple
relations between MUB and SIC quantization schemes exist in all
prime dimensions (MUBs and SIC-POVMs are compared also in
\cite{albouy,appleby-MUBs}).

\begin{table}
\caption{\label{table1} Sign function $S(k; a,\alpha)$.}
\begin{indented}
\item[]\begin{tabular}{@{}lcccccc} \br
& \centre{6}{$a\alpha$}\\
\ns
& \crule{6}\\
$k$ & 00 & 01 & 10 & 11 & 20 & 21\\
\mr
1 & $+$ & $-$ & $+$ & $-$ & $+$ & $-$\\
2 & $+$ & $-$ & $-$ & $+$ & $-$ & $+$\\
3 & $-$ & $+$ & $+$ & $-$ & $-$ & $+$\\
4 & $-$ & $+$ & $-$ & $+$ & $+$ & $-$\\
\br
\end{tabular}
\end{indented}
\end{table}

\section{\label{section-S-G}MUB-tomography and Stern-Gerlach measurements}
On passing a beam of spin-$j$ particles in a state $\hat{\rho}$
through Stern-Gerlach apparatus oriented along $z$-axis, we are
able to measure probabilities to find particles in each splitted
beam, i.e. in the state $|jm\rangle$, where $m=-j,\ldots,j$ is the
spin projection on $z$-axis. States $\{|jm\rangle\}_{m=-j}^{j}$
form the first basis in $d$-dimensional Hilbert space with
$d=2j+1$. Suppose a magnetic field ${\bf B}_a$ is applied to spin
particles before they are passed through the Stern-Gerlach
magnetic field gradient. This results in a unitary transformation
of the initial state $\hat{\rho} \rightarrow \hat{u}_a^{\dag}
\hat{\rho} \hat{u}_a$. The probabilities of outcomes read
$p_{a}(m) = \langle jm | \hat{u}_a^{\dag} \hat{\rho} \hat{u}_a |
jm \rangle$. On the other hand, $p_{a}(m) \equiv p_{a\alpha} =
\langle a\alpha | \hat{\rho} | a\alpha \rangle$, where
$|a\alpha\rangle = \hat{u}_a | j,m=\alpha - j \rangle$,
$\alpha=0,\ldots,2j$. States $\{|a\alpha\rangle\}_{\alpha=0}^{2j}$
form a new basis in Hilbert space, with parameter $a$ being a
label of this basis. Thus, applying different magnetic fields
${\bf B}_a$, $a=0,\ldots,2j+1$, we obtain a set of $2j+2$ bases
$\{|a\alpha\rangle\}_{\alpha=0}^{2j}$. If unitary transformations
$\hat{u}_a$ satisfy an additional condition $\left| \langle jm |
\hat{u}_a^{\dag} \hat{u}_b | jm' \rangle \right|^2 =
(2j+1)^{-1}(1-\delta_{a,b}) + \delta_{a,b}\delta_{m,m'}$, then the
constructed bases are MUBs.

However, application of constant magnetic fields $\{{\bf
B}_a\}_{a=0}^{2j+1}$ gives rise to unitary transformations
$\{\hat{u}_a\}_{a=0}^{2j+1}$ of the group $SU(2)$. Conversely, the
MUB-condition (\ref{MUBs}) can be only met when $u_a \in SU(N)$
with $N=2j+1$. As a result, MUBs can be constructed via the
conventional Stern-Gerlach technique for qubits ($j=\frac{1}{2}$)
only. For higher spins, the minimal necessary number of unitary
transformations $\hat{u}_a \in SU(2)$ is known to be
$4j+1$~\cite{serg-inverse-spin} which is greater than $2j+2$.

\section{\label{conclusions}Conclusions}
To conclude we summarize the main results of the paper.

Starting from peculiarities of mutually unbiased bases, we show
that whenever MUBs exist they can be used in quantum state
tomography. Then we develop the MUB-tomographic-probability
representation of quantum states by considering MUB-projectors
within the framework of a star-product scheme. Dequantizer and
quantizer of MUB star-product scheme are shown to be easily
expressed through MUB-projectors. This takes place due to a high
symmetry of MUBs. For the constructed MUB quantization scheme,
ordinary and dual star-product kernels are calculated and
expressed through triple product of MUB-projectors. Employing the
specific rules of the star-product kernel, we derive a new
relation on triple- and four-products. Applying the MUB star
product scheme to a commutator of MUB-projectors, we reveal the
Lie algebraic structure of MUB-projectors and find structure
constants. The obtained results can be used both in seeking and
classification of MUBs in higher dimensions.

Example of qubits is considered in detail. In particular, an
explicit form of all star-product characteristics is obtained.
Intertwining kernels between MUB and SIC star-product schemes are
found. These kernels can be used in order to find SIC-POVMs
whenever MUBs exist, for instance, in all prime dimensions.
Finally, an implementation of MUB-tomography via Stern-Gerlach
apparatus is discussed. The conventional experiment turns out to
be appropriate for MUB-tomography of qubits.

\ack The authors thank the Russian Foundation for Basic Research
for partial support under Project Nos. 09-02-00142 and
10-02-00312. SNF thanks the Ministry of Education and Science of
the Russian Federation and the Federal Education Agency for
support under Project No. 2.1.1/5909  and Contract No. $\Pi$558.
SNF is grateful to the Organizers of the 17th Central European
Workshop on Quantum Optics (St. Andrews, Scotland, UK, June 6-11,
2010) for invitation and kind hospitality. SNF would like to
express his gratitude to the Organizing Committee of the
Conference and especially to Dr. Natalia Korolkova and Dr. Irina
Leonhardt for financial support. SNF thanks the Russian Foundation
for Basic Research for travel grant No. 10-02-09331.

\end{document}